# Title: Earth's chondritic Th/U: negligible fractionation during accretion, core formation, and crust - mantle differentiation

**Authors:** Scott A. Wipperfurth[1†*], Meng Guo[1,2†], Ondřej Šrámek[3], William F. McDonough[1,4*]

**Affiliations:**

[1]Department of Geology, University of Maryland, College Park, MD 20742, USA

[2]Institute of Crustal dynamics, China Earthquake Administration, Beijing 100085, China

[3]Department of Geophysics, Faculty of Mathematics and Physics, Charles University, Prague, Czech Republic

[4]Research Center for Neutrino Science and Department of Earth Sciences, Tohoku University, Sendai 980-8578, Japan

[†]These authors contributed equally to this paper.

*Correspondence to: swipp@umd.edu and mcdonoug@umd.edu

**Abstract**: Radioactive decay of potassium (K), thorium (Th), and uranium (U) power the Earth's engine, with variations in $^{232}$Th/$^{238}$U recording planetary differentiation, atmospheric oxidation, and biologically mediated processes. We report several thousand $^{232}$Th/$^{238}$U ($\kappa$) and time-integrated Pb isotopic ($\kappa_{Pb}$) values and assess their ratios for the Earth, core, and silicate Earth. Complementary bulk silicate Earth domains (i.e., continental crust $\kappa_{Pb}^{CC} = 3.94\ ^{+0.20}_{-0.11}$ and modern mantle $\kappa_{Pb}^{MM} = 3.87\ ^{+0.15}_{-0.07}$, respectively) tightly bracket the solar system initial $\kappa_{Pb}^{SS} = 3.890 \pm 0.015$. These findings reveal the bulk silicate Earth's $\kappa_{Pb}^{BSE}$ is $3.90\ ^{+0.13}_{-0.07}$ (or Th/U = 3.77 for the mass ratio), which resolves a long-standing debate regarding the Earth's Th/U value. We performed a Monte Carlo simulation to calculate the $\kappa_{Pb}$ of the BSE and bulk Earth for a range of U concentrations in the core (from 0 to 10 ng/g). Comparison of our results with $\kappa_{Pb}^{SS}$ constrains the available U and Th budget in the core. Negligible Th/U fractionation accompanied accretion, core formation, and crust - mantle differentiation, and trivial amounts of these elements (0.07 ppb by weight, equivalent to 0.014 TW of radiogenic power) were added to the core and do not power the geodynamo.

1. *Introduction*

The Earth's engine is powered by an unknown proportion of primordial and radiogenic power (Šrámek et al., 2013; Jellinek and Jackson, 2015). Hence, we lack a gauge informing us as to how much and what proportion of fuel from either source remains in the Earth and what fraction is stored in the continental crust, mantle, and/or the core (Wohlers and Wood, 2015). This fuel powers volcanoes to erupt, the mantle to convect, tectonic plates to separate, and contributes to the generation of the protective magnetosphere that shields Earth's life from harmful cosmic rays.

Radioactive decay of K, Th, and U produces heat and contributes more than 99% of the non-primordial power to the Earth. The planet's primordial power comes from the accretion energy derived from assembling a uniform Earth and the gravitational energy release accompanying core formation. Given the Earth's size and mass, the accretion energy is of the order $10^{32}$ joules (or an equivalent temperature increase of the planet of tens of thousands of degrees), whereas the energy released by core–mantle differentiation is an order of magnitude smaller.

Defining the amount and distribution of the radioactive elements inside the Earth will exclude competing models of the meteoritic building blocks of planet construction, which resulted in its unique composition (Šrámek et al., 2013; Javoy and Kaminski, 2014; Wohlers and Wood, 2015; Jellinek and Jackson, 2015; Wohlers and Wood, 2017). It has been speculated that under certain conditions of planet formation a minor to significant fraction of these heat producing elements may have been sequestered into the Earth's core (Wohlers and Wood, 2015; Chidester et al., 2017; Blanchard et al., 2017). In addition, speculations abound on the existence of deep Earth reservoirs, which remain poorly sampled and mostly unaccounted for in the chemical descriptions of the Earth (Rizo et al., 2016; Jackson et al., 2017).

Here we present a global assessment of the relative abundance of Th and U in the Earth and its distribution — specifically between the crust, mantle, and core — and examine the degree to which fractionation of these elements has deviated from the primordial Th/U value seen in primitive meteorites (i.e., troilites from iron meteorites and chondrites). Our findings place into perspective this ratio and its importance in constraining chemical and biological processes that have occurred in the past.

Importantly, Th and U are refractory elements, those that condensed from the nebula at high temperatures (>1350 K), and are assumed to be accreted in chondritic proportions, with limited variability (circa ≤ ±10%) in their ratio. Both Th and U decay to separate isotopes of lead ($^{232}$Th to $^{208}$Pb and $^{238}$U to $^{206}$Pb) and thus one can evaluate the Earth's Th/U ratio using both the measured molar $^{232}$Th/$^{238}$U values (κ) and their time-integrated Pb isotopic values ($κ_{Pb}$) (Tatsumoto et al., 1973; Tatsumoto, 1978; Galer and O'Nions, 1985). The $\kappa_{Pb}$ ratio of the bulk silicate Earth's (BSE) reservoirs is derived by calculating the $^{208*}Pb/^{206*}Pb$ ratio after subtraction of the BSE initial $^{208}Pb/^{206}Pb$ value and time integrating the contributions from the two decay chains.

Recent estimates for the chondritic/solar system value of Th/U were reported from a Pb-Pb isotope array using troilites from iron meteorites ($\kappa_{Pb}^{SS}$= 3.890 ± 0.015 (Blichert-Toft et al., 2010)) and from seven dissolutions of the carbonaceous chondrite, Allende (Th/U$_{(mass\ ratio)}$ 3.77 ± 0.07 (Pourmand and Dauphas, 2010)). We recalculated the $\kappa_{Pb}^{SS}$ value reported in (Blichert-Toft et al., 2010)) using their reported initial $^{208}Pb^*/^{206}Pb^*$ value of 0.9572 ± 0.0038 and the following inputs: age of the Earth (4.568 x $10^9$ years), and the decay constants, $\lambda_{238U}$ = 1.5514 x $10^{-10}$ ($t_{1/2}$ = 4.468 x $10^9$ yr) and $\lambda_{232Th}$ = 4.9511 x $10^{-10}$ ($t_{1/2}$ = 14.0 x $10^9$ yr). Using these inputs and the $^{238}U$ natural molar isotopic fraction of 0.992742, we calculate a Th/U$_{(molar)}$ = 3.861 ± 0.016 and a Th/U$_{(mass)}$ = 3.764 ± 0.016 for this troilite dataset. Thus, the troilite and chondritic results are in full agreement and establish Earth's Th/U$_{(mass)}$ value, assuming chondritic proportions of refractory lithophile elements (McDonough and Sun, 1995).

We report a new global compilation of $\kappa$ and $\kappa_{Pb}$ values for oceanic and continental rocks, with the former samples serving as a proxy for the modern mantle (MM) and the latter for the bulk of the continental crust (CC). Following this we use the solar system/chondritic value of $\kappa_{Pb}$ and the data for the BSE to draw conclusions about the negligible contribution of the Earth's core in these heat producing elements' balance.

2. *Methods*

The continental crust and the modern mantle are understood as complementary reservoirs of the BSE. Their $\kappa$ and $\kappa_{Pb}$ values have been previously examined (Galer and O'Nions, 1985; Elliott et al., 1999; Paul et al., 2003; Andersen et al., 2015; Castillo, 2016; Kumari et al., 2016). In combination, the reservoirs' $\kappa$ ratios combine to the BSE $\kappa$ following these mass balance relationships:

$$M^{BSE} = M^{MM} + M^{CC} \quad [1]$$

$$M^{BSE} a_{Th}^{BSE} = M^{MM} a_{Th}^{MM} + M^{CC} a_{Th}^{CC} \quad [2]$$

$$M^{BSE} a_{U}^{BSE} = M^{MM} a_{U}^{MM} + M^{CC} a_{U}^{CC} \quad [3]$$

where equations (2) and (3) can be combined into

$$\left(\frac{a_{Th}}{a_U}\right)^{BSE} = \frac{M^{MM} a_{Th}^{MM} + M^{CC} a_{Th}^{CC}}{M^{BSE} a_U^{BSE}} = \left(\frac{a_{Th}}{a_U}\right)^{MM} \frac{M^{MM} a_U^{MM}}{M^{BSE} a_U^{BSE}} + \left(\frac{a_{Th}}{a_U}\right)^{CC} \frac{M^{CC} a_U^{CC}}{M^{BSE} a_U^{BSE}}$$

$$= \left(\frac{a_{Th}}{a_U}\right)^{MM} \frac{m_U^{MM}}{m_U^{BSE}} + \left(\frac{a_{Th}}{a_U}\right)^{CC} \frac{m_U^{CC}}{m_U^{BSE}} \quad [4]$$

and where *M*, $m_U$, and $m_{Th}$ represent the mass of rock, U, and Th in the reservoir of interest, *a*

= the abundance of U or Th (mass fraction), *MM* = modern mantle, and *CC* = continental crust. Multiplication of the above equations by the ratio in atomic masses of Th and U and by the molar fraction of $^{238}$U relative to U will yield κ. Thus, calculation of the κ and $κ_{Pb}$ of the BSE requires weighting the κ and $κ_{Pb}$ of the modern mantle or continental crust by only the U mass within each reservoir.

3. *Data*

Some 150,000 measurements acquired through the *EarthChem* (www.earthchem.org) repository were statistically assessed and are provided as an electronic supplement. Select measures, including the arithmetic mean, geometric mean, and median are reported in Table 1 for oceanic basalts and continental igneous, sedimentary, and metamorphic rocks. Moreover, this analysis provides the full assessment of the data distribution and uncertainties (Table 1 and Supplementary Information). Sample location maps and histograms of κ and $κ_{Pb}$ are provided in Figures 1, 2 and 3. Data trends versus $SiO_2$, along with regression analyses are reported for continental rocks in Table 1 and Figure S1. The electronic supplement contains the following: (i) detail the κ and $κ_{Pb}$ definitions and analysis; (ii) description of the κ versus $SiO_2$ regression for continental rocks; (iii) a description of the method of assessing the crustal composition using crustal types; (iv) details of the mathematical formulation; and (v) additional figures and tables.

Significant analytical advances over the last few decades allow for a reassessment of the κ value for fresh, unaltered mid-ocean ridge basalt (MORB). Table 1 reports the arithmetic mean, median, and geometric mean for 2,558 high precision κ values for MORB, many of which have been recently determined by laser ablation measurements (Arevalo and McDonough, 2010; Gale et al., 2013). The present survey provides an improved estimate of the central value and 68% confidence limits for κ and $κ_{Pb}$ of MORB (Figure 1 and 3). The similarity of the arithmetic mean, median, and geometric mean for $κ^{MORB}$ of ~3.1 ± 0.7 are consistent with a gaussian data distribution. This value and its uncertainty is comparable, at the low end of the limit, to earlier estimates of ~2.5 (Galer and O'Nions, 1985; Elliott et al., 1999; Paul et al., 2003). The $κ_{Pb}^{MORB}$ of 3.84 ± 0.09 for MORB is identical to earlier estimates and overlaps with that of the solar system initial (3.890 ± 0.015). The homogeneity of the MORB $κ_{Pb}$ stands in contrast to the greater variability seen in κ (Figure 3).

Similarly, the statistics for κ and $κ_{Pb}$ of ocean island basalts (OIB) have been enhanced based on many thousand measurements and improved data quality. The $κ^{OIB}$ value ($3.67\ ^{+0.99}_{-0.63}$) is more variable than that for MORB, whereas the $κ_{Pb}^{OIB}$ value ($3.87\ ^{+0.16}_{-0.07}$) is comparable to that for MORB and again overlaps with that of the solar system value (3.890 ± 0.015). We treat OIB and MORB samples collectively as representing the Modern Mantle (MM), with the former sampling a dominantly incompatible element enriched source and the latter an incompatible element depleted source. The relative mass fractions of these two mantle domains and their concentrations are not well constrained, but estimates place the OIB source as representing about 20% the mass of the Modern Mantle (Arevalo et al., 2013). In any case, thanks to the similarity of the $\kappa_{Pb}$ values for MORB and OIB, our subsequent analysis is largely independent of the reservoirs' relative sizes. Simple combining of the OIB and MORB datasets yield a modern mantle $\kappa^{MM} = 3.54\,^{+0.96}_{-0.69}$ and $\kappa_{Pb}^{MM} = 3.87\,^{+0.15}_{-0.07}$. Alternatively, assuming mantle mass fractions and concentrations of OIB and MORB from (Arevalo et al., 2013) yields a $^{weighted}\kappa^{MM} = 3.46\,^{+0.72}_{-0.57}$ and $^{weighted}\kappa_{Pb}^{MM} = 3.86\,^{+0.09}_{-0.06}$. The similarity of the weighted and unweighted $\kappa_{Pb}^{MM}$ is a result of the enrichment in U and Th of the OIB source relative to the MORB source. We use the unweighted values in later calculations.

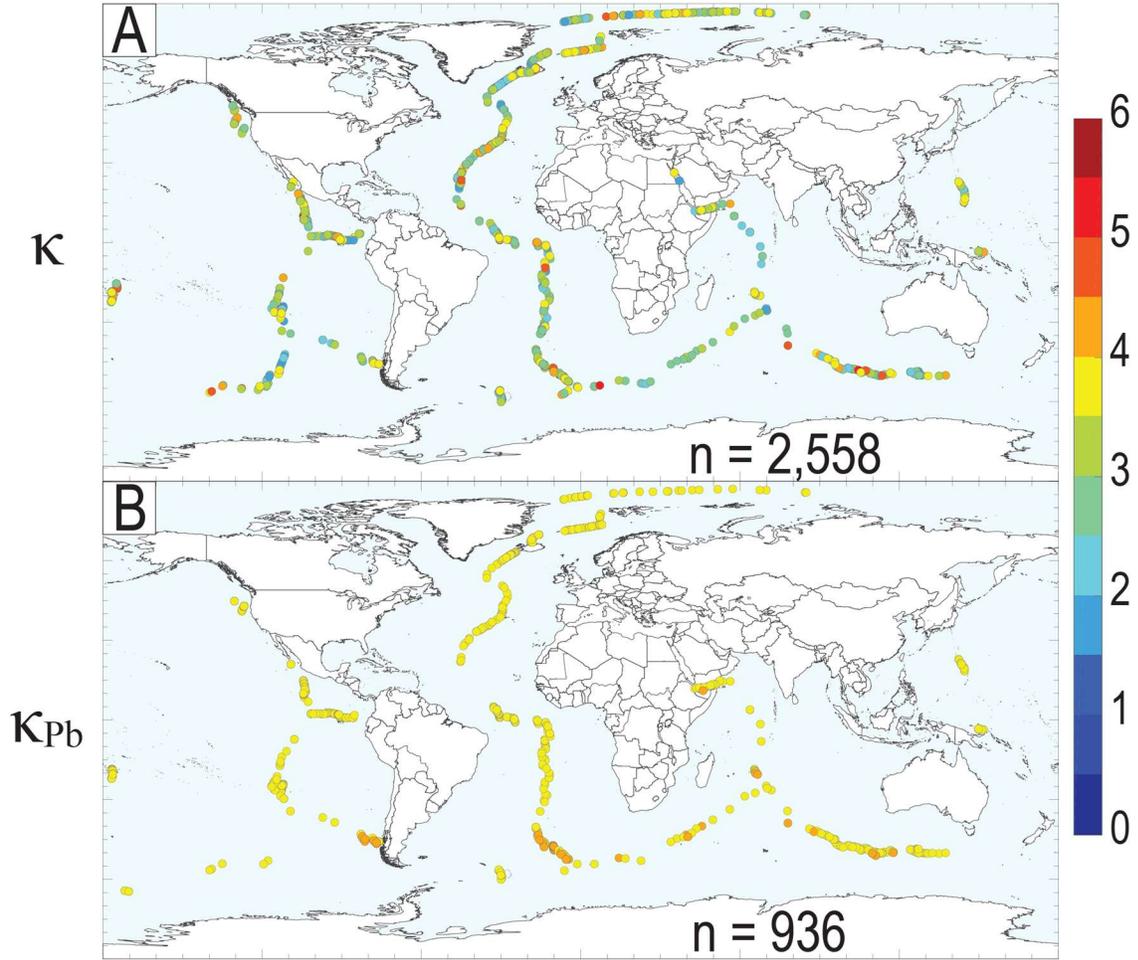

**Fig. 1**: Distribution of samples used in the calculation of κ (A) and $\kappa_{Pb}$ (B) for MORB. Circle colors indicate κ and $\kappa_{Pb}$ value. The number of samples comprising each dataset are reported on the bottom right of each map.

The continental crust is the remaining known BSE reservoir for Th and U and its contribution is estimated to contain some ~30% to 50% of the inventory of these heat producing elements (Huang et al., 2013; Rudnick and Gao, 2014). Assessing the κ and $κ_{Pb}$ values for the bulk of the continental crust presents a challenge, given different rock types from which to select, sampling biases, and weighting fractions for the values for the upper, middle, and lower crust. Therefore, to evaluate κ values in igneous, metamorphic, and sedimentary rocks we used a best fit line between κ and $SiO_2$ to calculate the average κ value at 60 wt.% $SiO_2$ (Table 1 and Figure S1). Results of these and an additional method agree with statistical measures for the unweighted datasets, particularly for $κ_{Pb}$ (Table 1 and Table S3). We use the unweighted values in later calculations.

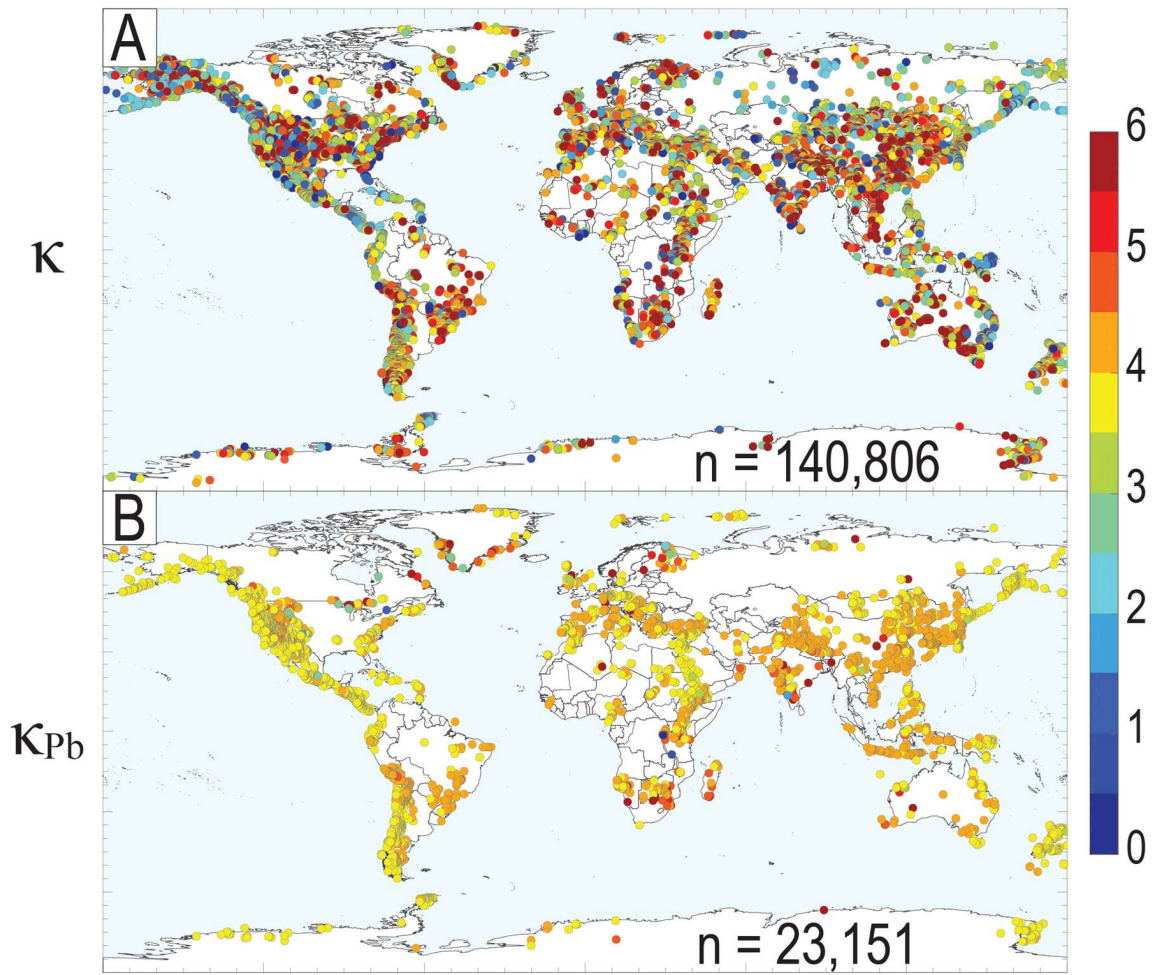

**Fig. 2**: Distribution of samples used in the calculation of κ (A) and $κ_{Pb}$ (B) for continental crust (igneous, sedimentary, and metamorphic datasets). Circle colors indicate κ and $κ_{Pb}$ value. The number of samples comprising each dataset are reported on the bottom right of each map.

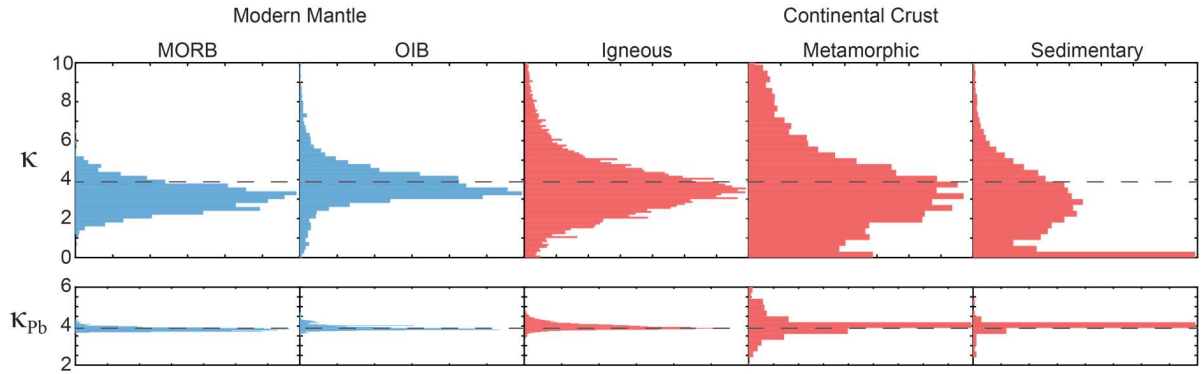

**Fig. 3.** Histograms of κ and $κ_{Pb}$ for MORB, OIB, and continental igneous, metamorphic, and sedimentary rocks used in this study. The y-axis shows κ and $κ_{Pb}$ values at the same scale and the x-axis shows frequency of sample per bin normalized to total number of samples. The dashed dark-grey horizontal line in each plot is the solar system initial value (Blichert-Toft et al., 2010).

Marked differences are found between the gaussian fit (arithmetic mean and standard deviation), log-normal fit (geometric mean and corresponding asymmetrical uncertainty), and the median with 68% confidence limits, for κ and Th and U abundances for igneous, sedimentary, and metamorphic rocks (Table 1 and Table S3), whereas $κ_{Pb}$ values are nearly identical in these different datasets (Table 1). The uniformly lower κ values in sedimentary rocks (Figure 3, including the peak for U rich sediments with low Th/U values), reflect the mobility of U presently at oxidized surface conditions; these samples likely add a small contribution (~10%) to the overall continental crustal signal (Wilkinson et al., 2009). The median, geometric mean, and 60 wt.% $SiO_2$ estimates of κ values (~3.6) for igneous and metamorphic rocks are low when compared to conventional estimates (i.e., Th/U > ~4 for the continental crust), but their wide confidence limits at 68% and markedly asymmetric distribution indicate a positively skewed population (Table 1 and Figure 3). The median and geometric mean κ values of mantle and crustal datasets are not complimentary relative to the chondritic reference frame, albeit are within uncertainty. The arithmetic mean, geometric mean, and median of $κ_{Pb}$ for continental igneous and metamorphic rocks overlap and are slightly super-chondritic. The median $κ_{Pb}$ values show a complementary relationship with the OIB and MORB values (Table 1).

The continental crust $κ_{Pb}^{CC}$ of 3.95 $^{+0.20}_{-0.11}$ (combined igneous and metamorphic data) overlaps with previously reported values calculated from Pb isotopic compositions for the continents (Rudnick and Goldstein, 1990), although some of these higher model estimates (Zartman and Doe, 1981) only marginally overlap with the data presented here. Also, the $κ_{Pb}^{sed}$ value of 4.00 $^{+0.11}_{-0.14}$ for sedimentary rocks agrees with the estimate of 4.04 based on the Pb isotopic composition of the upper crust (Millot et al., 2004). (Blichert-Toft et al., 2010)

Concentration data for Th and U in rocks from the continental crust and modern mantle are provided in Table S1. In all cases the estimated concentrations based on the arithmetic mean, geometric mean, and median reveal that the latter two are comparable and lower than the arithmetic mean, indicating a non-gaussian, positively skewed data distribution in all geological settings. The arithmetic mean of Th and U abundances for MORB are comparable to earlier estimates (Arevalo and McDonough, 2010; Gale et al., 2013). Estimates of Th and U abundance in the continental crust, based on arithmetic mean, are higher than existing estimates for the upper crust and reflect a strong biasing of the data by outliers with high concentrations. The median and log-normal estimates of Th and U abundances in igneous and metamorphic rocks are comparable with global estimates for the bulk continental crust (Rudnick and Gao, 2014); based only on the median abundances the Th/U is ~3.6 to 3.8, comparable to the weighted median κ value for these rocks.

4. *Calculating the Earth's core contribution*

We also examine the role of the Earth's core in establishing the budget of these elements. To date, experimental studies conducted at conditions ranging from low to high pressure and high temperature have examined the partitioning of Th and U between metal (and metal-sulfide) and silicate and document a marked difference in the behavior of these two elements, with U, and not Th, being weakly partitioned into the metal (Chidester et al., 2017; Blanchard et al., 2017). These same studies either did not include Th in their experiments or report $D_U$ partition coefficients and below limits of detection values for $D_{Th}$ (Chidester et al., 2017). The Earth's core is recognized as having a mix of light elements that account for its density deficit relative to iron, with a limited contribution from the volatile element sulfur (McDonough, 2017). In a sulfur saturated, non-peridotitic set of experiments, $D_{Th}$ was found to be ~$0.1 D_U$ (Wohlers and Wood, 2017).

The mass balance of the bulk Earth (⊕) is:

$$m_U^{\oplus} = m_U^{CC} + m_U^{MM} + m_U^{core} \quad [5]$$

The core/mantle (i.e., metal/silicate) partition coefficient is defined as $D_{Th} \equiv a_{Th}^{core}/a_{Th}^{BSE}$. We further define

$$D_\kappa \equiv \frac{D_{Th}}{D_U} = \left(\frac{Th}{U}\right)^{Core}\left(\frac{U}{Th}\right)^{BSE} = \frac{\kappa^{core}}{\kappa^{BSE}} \quad [6]$$

to express

$$\kappa_{Pb}^{core} = D_\kappa \kappa_{Pb}^{BSE} = D_\kappa\left(\kappa_{Pb}^{CC}\frac{m_U^{CC}}{m_U^{BSE}} + \kappa_{Pb}^{MM}\frac{m_U^{MM}}{m_U^{BSE}}\right) \quad [7]$$

As U is removed from the BSE (specifically the mantle) and added to the core, the mass balance between the mantle and crust (equation [8]) changes, as does $\kappa_{Pb}^{BSE}$ and $\kappa_{Pb}^{core}$ (equation [7]). We use $D_\kappa = 0.1$ as an upper limit during metal-silicate segregation (Wohlers and Wood, 2017). Finally, the bulk Earth $\kappa_{Pb}^{\oplus}$ is calculated as

$$\kappa_{Pb}^{\oplus} = \kappa_{Pb}^{CC} \frac{m_U^{CC}}{m_U^{\oplus}} + \kappa_{Pb}^{MM} \frac{m_U^{MM}}{m_U^{\oplus}} + \kappa_{Pb}^{core} \frac{m_U^{core}}{m_U^{\oplus}} \quad [8]$$

Using the data (Table 1 and S1) and its distribution (Figure 3), we performed a Monte Carlo simulation to calculate the $\kappa_{Pb}$ of the BSE and bulk Earth for a range of U concentrations in the core (from 0 to 10 ng/g) and a bulk Earth U mass from (McDonough and Sun, 1995). The lower limit assumption of $m_U^{core} = 0$ puts all the U in the BSE (equation [5]). As U is added to the core it is removed from the mantle U mass budget. We adopt weights based on the U mass fraction in the continental crust and mantle (equation [4]). κ and $\kappa_{Pb}$ were randomly sampled $10^5$ times from the available data for each reservoir, avoiding assumptions on distribution shape. We re-sample the distributions in cases where the mass of U in the combined crust and core reservoirs result in a larger U mass than permitted by the assumed bulk Earth model. This re-sampling removed a subset of crustal data with high U abundances, and resulted in a $\kappa^{CC} = 3.64^{+1.49}_{-1.37}$, similar to results reported in Table 1. We also observe a lack of correlation between U or Th abundance, and κ.

Based on an assumption of $U^{core}$ = 0 ng/g, the $\kappa_{Pb}^{BSE} = 3.895^{+0.130}_{-0.075}$ (equivalent to Th/U$_{BSE}$ (mass ratio) = 3.769 ± 0.126). Figure 4 shows the modeled results. The intersection of the median $\kappa_{Pb}^{\oplus}$ with the median $\kappa_{Pb}^{SS}$ yields a U abundance in the core of 0.07 ng/g. Considering the 68% and 95% confidence limits (c.l), the maximum allowable abundance of U in the core is 1.8 ng/g and 4.6 ng/g, respectively (Figure S3). The heat production in the core from U and Th (calculated using $\kappa_{Pb}^{core}$) is 0.014 TW, 0.39 TW, and 0.95 TW, respectively for the median, 68% c.l., and 95% c.l. intercepts.

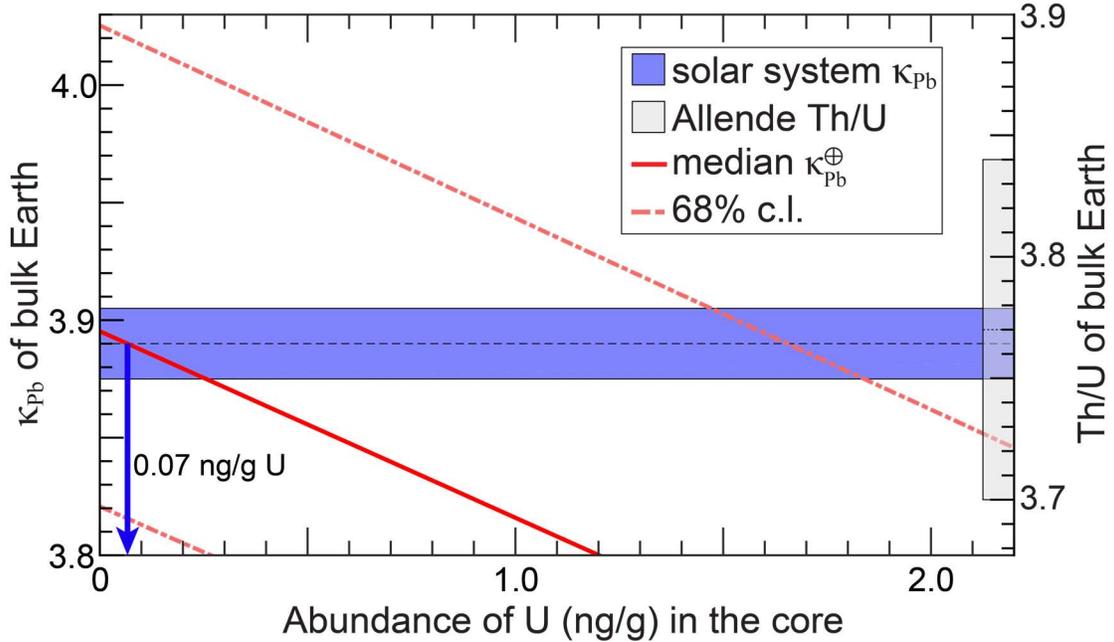

**Fig. 4.** Modeled U abundance (in ng/g) in the core vs. $\kappa_{Pb}$ of the the bulk Earth (left y-axis) and Th/U (mass ratio) of the bulk Earth (right y-axis). The median (red solid) and 68% confidence limit (c.l.; red dashed) are plotted against the solar-system $\kappa_{Pb}$ (blue shaded horizontal region with 1-sigma bounds; Blichert-Toft et al., 2010) and Allende meteorite (grey shaded region along right side with 1-sigma bounds and arbitrary x-axis size; Pourmand and Dauphas, 2010). The intersection of the median $\kappa_{Pb}$ of the bulk Earth and solar system is marked by a solid vertical blue line (corresponding to 0.07 ng/g).

5. *Discussion*

The results of this study find that the $\kappa_{Pb}^{MM} = 3.87^{+0.15}_{-0.07}$ and $\kappa_{Pb}^{CC} = 3.94^{+0.20}_{-0.11}$ are statistically indistinguishable, $\kappa_{Pb}^{MM} \approx \kappa_{Pb}^{CC}$, and therefore a mass weighted estimate for the $\kappa_{Pb}^{BSE} = 3.90^{+0.13}_{-0.07}$ is comparable to the solar system/chondritic value (Pourmand and Dauphas, 2010; Blichert-Toft et al., 2010). These findings resolve a long-standing debate on the BSE estimate of its Th/U value (Allègre et al., 1986; Allègre et al., 1988; Rocholl and Jochum, 1993; Javoy and Kaminski, 2014).

Second, negligible Th/U fractionation accompanied crust - mantle differentiation. The 0.9% difference between the median $\kappa_{Pb}^{CC}$ and $\kappa_{Pb}^{MM}$ reveals that $U^{6+}$ recycling back into the mantle has either been a relatively recent process or that limited recycling followed atmospheric oxygenation at 2.4 Ga and evolved slowly with time. These finding are strikingly inconsistent with claims of widespread pollution of the upper mantle with recycled uranium (Andersen et al., 2015) and claims of biologically driven fractionation leaving its imprint on the mantle (Sleep et al., 2013).

Third, negligible Th/U fractionation accompanied accretion and core formation, and therefore trivial amounts of U (<4.6 ng/g at 95% c.l., Figure 4 and S3) and proportionally miniscule amounts of Th have been sequestered into the Earth's core, and thus U and Th play little to no role in powering the geodynamo (Nimmo, 2015). A maximum of 0.95 TW is available from U and Th in the core at 95% confidence limits. Likewise, this finding falsifies the hypothesis of a natural nuclear reactor in the Earth's core (Hollenbach and Herndon, 2001).

Using the findings reported here, we recommend that the uncertainty analyses reported in (McDonough and Sun, 1995) be updated to ±10% for the abundances of Th and U (following the uncertainty of $\kappa_{Pb}^{BSE}$). This update sets the uncertainty values for these elements as being equal to that for the rest of the refractory lithophile elements.

The precision reported here for the global Th/U value validates and updates the assumed fixed ratio to Th/U (mass ratio) = 3.8 for use in geoneutrino detection studies (Gando et al., 2013; Agostini et al., 2015), particularly given the crust and mantle domains share similar values. That said, however, differences in this ratio may be found in crustal regions beneath individual detectors. An independent measure of this value by particle physicists is welcomed and will be a critical test of our global interpretation.

There remains debate regarding the absolute abundances of U and Th in the BSE, with estimates varying by a factor of three (Šrámek et al., 2013). The agreement of $\kappa_{Pb}^{CC}$ and $\kappa_{Pb}^{MM}$ and the results for the abundances of U and Th in the continental crust are combined with an assumption of a BSE model (McDonough and Sun, 1995) to show that the power balance for these elements is 65% in the mantle and 35% in the crust, a conclusion consistent with geoneutrino results (Gando et al., 2013; Agostini et al., 2015). These results highlight the significance of the 13 TW of power driving mantle convection. Our conclusions, nonetheless, are independent of BSE model chosen, given the near equivalence of $\kappa_{Pb}^{CC}$ and $\kappa_{Pb}^{MM}$. Moreover, our findings emphasize the need for further understanding the energy budget of the core.

**Acknowledgments:** The data reported in this paper are tabulated in the Supplementary Materials available online. Support for this study was provided by The U.S.-China Fulbright Program (to M.G.), NSF grant EAR1650365 (to W.F.M.), and Czech Science Foundation grant GAČR 17-01464S (to O.Š.).

|  | | κ | | | | | $κ_{Pb}$ | | | |
| --- | --- | --- | --- | --- | --- | --- | --- | --- | --- | --- |
| Reservoir | # of data | Mean | Geometric mean | Median | SiO$_2$ at 60 wt% | # of data | Mean | Geometric mean | Median | SiO$_2$ at 60 wt% |
| Modern Mantle | | | | | | | | | | |
| MORB | 2,558 | 3.14 | 3.05 | $3.12^{+0.72}_{-0.71}$ | | 936 | 3.84 | 3.84 | $3.84^{+0.09}_{-0.09}$ | |
| OIB | 10,599 | 4.07 | 3.66 | $3.67^{+0.99}_{-0.63}$ | | 6,576 | 3.91 | 3.90 | $3.87^{+0.16}_{-0.07}$ | |
| Continental Crust | | | | | | | | | | |
| igneous | 120,836 | 4.73 | 3.42 | $3.56^{+1.60}_{-1.29}$ | $3.61^{+1.0}_{-0.6}$ | 22,318 | 3.99 | 3.97 | $3.95^{+0.19}_{-0.11}$ | $3.95^{+0.07}_{-0.06}$ |
| metamorphic | 8,287 | 5.37 | 3.10 | $3.61^{+3.26}_{-2.00}$ | $3.57^{+1.8}_{-1.2}$ | 664 | 4.12 | 4.05 | $3.99^{+0.40}_{-0.28}$ | $3.99^{+0.3}_{-0.2}$ |
| sedimentary | 11,682 | 3.09 | 1.52 | $2.61^{+1.87}_{-2.16}$ | $2.60^{+0.9}_{-0.7}$ | 169 | 3.99 | 3.97 | $4.00^{+0.11}_{-0.10}$ | $4.00^{+0.08}_{-0.05}$ |

**Table 1.** Summary of $κ$ and $κ_{Pb}$ for MORB, OIB, and continental crust (igneous, metamorphic, and sedimentary datasets. 68% confidence limits are reported alongside the median value for each dataset and method. Weighting by SiO$_2$ at 60 wt% is also reported for the continental crust (see text and supplementary materials for details).